\documentclass[10pt,letter,final]{IEEEtran}
\usepackage{amsthm,amssymb,graphicx,multirow,amsmath,color,amsfonts}
\usepackage[update,prepend]{epstopdf}%
\usepackage[noadjust]{cite}
\usepackage{tikz}
\usetikzlibrary{arrows,calc}		
\usepackage{bbm} 
\usepackage{pdfpages}
\usepackage{tabulary}
\usepackage{multirow}
\usepackage{comment}
\usepackage{url}
\usepackage{dsfont}
\usepackage{pgf,tikz}
\usepackage{hyperref}
\usepackage{bigints}
\usepackage{mathtools}
\usepackage[english]{babel}
\usepackage[autostyle]{csquotes}
\usepackage{textcomp}
\usepackage{varwidth}
\usepackage[normalem]{ulem}

\newcommand{\pgfl}{\mathtt{PGFL}}
\newcommand{\cdf}{\mathtt{CDF}}

\newcommand\numberthis{\addtocounter{equation}{1}\tag{\theequation}}

\makeatletter
\renewcommand{\env@cases}[1][@{}l@{\quad}l@{}]{%
  \let\@ifnextchar\new@ifnextchar
  \left\lbrace
  \def\arraystretch{1.2}%
  \array{#1}%
}
\makeatother

\def\nb0{{\mathbf{0}}}
\def\nb1{{\mathbf{1}}}

\def\nbbP{{\mathbb{P}}}

\newtheorem{thm}{Theorem}
\newtheorem{definition}{Definition}

\def\E{\mathbb{E}}

\def\P{\mathbb{P}}

\def\R{\mathbb{R}}

\def\sir{\mathtt{SIR}}
\def\pcf{\mathtt{pcf}}

\def\calI{\mathcal{I}}
\def\calB{\mathcal{B}}

\def\calL{\mathcal{L}}

\newcommand{\dP}[1]{\mathbb{P}\left[#1\right]}

\allowdisplaybreaks 

\begin{document}
\graphicspath{{./Figures/}}
\title{ Downlink Analysis for the Typical Cell\\ in Poisson Cellular Networks}
\author{
Praful D. Mankar, Priyabrata Parida, Harpreet S. Dhillon, Martin Haenggi
\thanks{P. D. Mankar, P. Parida and H. S. Dhillon are with Wireless@VT, Department of ECE, Virginia Tech, Blacksburg, VA. Email: \{prafuldm, pparida, hdhillon\}@vt.edu. M. Haenggi is with the Department of Electrical Engineering, University of Notre Dame, Notre Dame, 
IN. Email: mhaenggi@nd.edu.
This work was supported by the US NSF under Grant ECCS-1731711.
}
}

\maketitle
\begin{abstract}
Owing to its unparalleled tractability, the Poisson point process (PPP) has emerged as a popular model for the analysis of cellular networks. Considering a stationary point process of users, which is independent of the base station (BS) point process, it is well known that the typical user does not lie in the typical cell and thus it may not truly represent the typical cell performance. Inspired by this observation, we present a construction that allows a direct characterization of the downlink performance of the typical cell. For this, we present an exact downlink analysis for the 1-D case and a remarkably accurate approximation for the 2-D case. Several useful insights about the differences and similarities in the two viewpoints (typical user vs. typical cell) are also provided.
\end{abstract}\vspace{-.15cm}
\vspace{-.35cm}
\section{Introduction}
\label{sec:Introduction}
The previous decade has witnessed a significant growth in research efforts related to the modeling and analysis of cellular networks using stochastic geometry. A vast majority of these works, e.g., \cite{AndBacJ2011,DhiGanJ2012}, rely on the homogeneous PPP model for the BS locations. The user locations are then modeled as a stationary point process that is assumed to be independent of the BS process. Given the stationarity and independence of the user point process, the concept of coverage of the typical user and coverage of an arbitrary fixed location are identical. As a result, one does not need to explicitly consider Palm conditioning on the user point process and the analysis can just focus on the origin as a location of the typical user. However, it is well known that the origin falls in a Poisson-Voronoi (PV) cell that is bigger on an average than the typical cell \cite{BacBla2009}, called the {\em Crofton cell}. Therefore, this approach does not characterize the performance of the typical cell, which is the main focus of this letter. 

One way of characterizing the typical cell performance is to consider a user distribution model that places a single user distributed uniformly at random in each cell independently of the other cells. This user process can be interpreted as the locations of the users scheduled in a given resource block. One can also argue that this point process is at least as meaningful as the one discussed above because practical cellular networks are dimensioned to ensure that the load of each cell is almost the same. Similar to~\cite{Haenggi2017}, we refer this user process as {\em Type I user process} and the aforementioned independent user process as  {\em Type II user process}. Given that the Crofton cell is statistically larger than the typical cell, it is easy to establish that both the desired signal power and the interference power observed at the typical user of the Type I process will (stochastically) dominate the corresponding quantities observed by the typical user of the Type II process. 
While the downlink analysis of the Type II user process is well understood, this letter deals with the downlink coverage analysis for the Type I user process.

\indent{\em Related Works:} The downlink analysis of cellular networks with the Type II user process involves using the {\em contact distribution} of the PPP to characterize the link distance, and using {\em Slivnyak's theorem} to argue that conditioned on the link distance, the  point process of interferers remains a PPP~\cite{AndBacJ2011}. While the idea of using the Type I user process is relatively recent, there are two noteworthy works in this direction. First and foremost is~\cite{Haenggi2017}, which defined this user process and used it for the uplink analysis. This idea was extended to the downlink case in \cite{Martin2017_Meta}, where the meta distribution of signal-to-interference ratio ($\sir$) is derived using an empirically obtained link distance distribution (for the Type I process) and approximating the point process of interferers as a homogeneous PPP beyond the link distance from the location of the typical user. In \cite{PraPriHar}, we derived the exact integral expression and a closed-form approximation for the serving link distance distribution for the Type I user process. Building on the insights obtained from \cite{Martin2017_Meta} and \cite{PraPriHar}, we provide an accurate downlink analysis for the Type I user process in this letter. 

\indent{\em Contributions:} The most important contribution of this letter is to demonstrate that the well-accepted way of defining the typical user by considering an independent and stationary point process of users is not the only way of analyzing cellular networks modeled as point processes. More importantly, this construction does not result in the typical cell performance. In order to highlight the finer differences between the two viewpoints, we first present the exact analysis of the Type I process for the 1-D case.  Leveraging the qualitative insights obtained from the 1-D case, we perform an approximate yet accurate analysis for the Type I user process in a 2-D cellular network. 
In particular, for the Type I process, we empirically show that the point process of interfering BSs given a distance $R_o$ between user and serving BS  exhibits  a clustering effect at distances slightly larger than $R_o$ that is not captured by a homogeneous PPP approximation (beyond $R_o$) as used in \cite{Martin2017_Meta}. Using this insight, we propose a dominant-interferer based approach in order to accurately approximate the point process of interferers. This approximation allows us to accurately evaluate the interference received by the user conditioned on the link distance, which subsequently provides a remarkably tight approximation for the 2-D case.
\section{System Model and Preliminaries}
\label{sec:System_Model}
We assume that the locations of BSs form a homogeneous PPP
$\Psi\equiv\{\textbf{x}_1,\textbf{x}_2,\dots\}$ of density $\lambda$ on $\R^d$ for $d\in\{1,2\}$.
The PV cell with the nucleus at $\mathbf{x}\in\Psi$ can be defined as
\begin{equation}
V_\mathbf{x}=\{\mathbf{y}\in\mathbb{R}^d: \|\mathbf{y}-\mathbf{x}\|\leq\|\mathbf{x'}-\mathbf{y}\|, ~\forall \mathbf{x'}\in\Psi\}.
\label{eq:PV_Cell_Definition}
\end{equation}
Since by Slivnyak's theorem \cite{Haenggi2013}, conditioning on a point is the same as adding a point to a PPP, we focus on the typical cell of the point process $\Psi\cup\{o\}$ at $o$, which is given by
\begin{equation}
V_o=\{\mathbf{y}\in\mathbb{R}^d: \|\mathbf{y}\|\leq\|\mathbf{x}-\mathbf{y}\|, ~\forall \mathbf{x}\in\Psi\}.
\end{equation}
Henceforth, we consider $\Phi=\Psi\cup\{o\}$. Further, let  $\tilde{V}_o$ be the cell of the PV tessellation of $\Psi$ containing the origin, called the Crofton cell.
Without loss of generality, the typical user from the Type II user process can be assumed to be located at the origin (see \cite{AndBacJ2011}) which means  it resides in the Crofton cell $\tilde{V}_o$.  
Now, we define Type I user point process as
\begin{equation}
\Omega\triangleq\{U(V_\mathbf{x}):\mathbf{x}\in\Phi\},
\label{eq:UserProcessTypeI}
\end{equation} 
where $U(A)$ is the point chosen uniformly at random from the set $A$ independently for different $A$. Note that the typical user from  the Type I user point process $\Omega$ represents a uniformly random point in the typical cell.

 By the above construction, the location of the typical user in the typical cell becomes $\mathbf{y}\sim U(V_o)$ and $\Psi$ becomes the point process of interfering BSs to the typical user at $\mathbf{y}\in V_o$.  
Let $R_o$ denote the {\em link distance}, i.e., the distance from the BS of $V_o$ (i.e., the origin) to the user at $\mathbf{y}$.
We consider the standard power law path loss model with exponent $\alpha>2$ for signal propagation. Further, assuming independent  Rayleigh fading, we model the small-scale fading gains $h_{\mathbf{x}}$ associated with the typical user and the BS at $\mathbf{x}\in\Phi$ as exponentially distributed random variables with unit
mean. We assume $\{h_{\mathbf{x}}\}$ are independent for all $\mathbf{x}\in\Phi$.
Thus, $\sir$ at the typical user located at $\mathbf{y}\in V_o$ in an interference-limited system is
\begin{equation}
\sir=\frac{h_{o}R_o^{-\alpha}}{\sum\limits_{\mathbf{x}\in\Psi}h_{\mathbf{x}}\|\mathbf{x}-\mathbf{y}\|^{-\alpha}}.
\end{equation}\vspace{-.3cm}
\begin{definition}
The coverage probability is the probability that the $\sir$ at the typical user is greater than a threshold $\tau$.
\end{definition}\vspace{-.2cm}
In the rest of this section, we briefly discuss the coverage probability of the Type II process.

By definition, the link distance of the typical user of the Type II user process is $\hat{R}_o=\|\mathbf{x}\|$ where $\mathbf{x}\in\Psi$ is the closest point to the origin. The cumulative distribution function ($\cdf$) of $\hat{R}_o$ (i.e., the {\em contact distribution}) is  $1-\exp(-\lambda\kappa_dr^d)$ \cite{Haenggi2013},  where $\kappa_d=1$ and $\kappa_d=\pi$ for $d=1$ and $d=2$, respectively.
The coverage probability of the Type II process in the $d$-dimensional Poisson cellular network is given by \cite{AndBacJ2011} 
\begin{align}
\mathtt{P_{II}^d}(\tau)\triangleq\nbbP[\sir>\tau]= &\left[{1+\tau^{\frac{d}{\alpha}}\int_{\tau^{-\frac{d}{\alpha}}}^{\infty} \frac{1 }{1 + u^{\frac{\alpha}{d}}} {\rm d}u}\right]^{-1}. \numberthis
\label{eq:CovP_Crf_d12}
\end{align}
Note that \cite{AndBacJ2011} is focused on the case of $d=2$ but the extension to the general $d$-dimensional case is straightforward. $\mathtt{P_{II}^d}(\tau)$ can also be interpreted as the fraction of the covered area.
\section{Coverage Analysis of Type I User Process}
\label{sec:CovP_TypCell}
In this section, we present the exact and an approximate (yet accurate) coverage analysis of the Type I user process for $d=1$ and $d=2$.
\vspace{-.45cm}
\subsection{Exact Coverage Analysis for $d=1$}
\label{sec:CovP_Analysis_1d}
We begin our discussion with the distribution of the serving link distance conditioned on the distances from the typical BS at the origin to the neighboring BSs (one from each side). Let $R_1$ and $R_2$ be the distances from the typical BS to these two neighboring BSs. Since $\Phi$ is a Poisson process on $\R$, $R_1$ and $R_2$ are i.i.d. exponential with mean $\lambda^{-1}$. The joint distribution of $R_1$ and $R_2$ conditioned on $R_1<R_2$ is 
\begin{align}\
\hspace{-.1cm}f_{R_1,R_2}(r_1, r_2) = 2 \lambda^2 \exp(-\lambda (r_1+r_2)), ~r_2\geq r_1 \geq 0. \numberthis
\label{eq:DIST_R1R2}
\end{align}
The serving link distance distribution for the user at $y\sim U(V_o)$, where $|y|=R_o$,  conditioned on $R_1$ and $R_2$ becomes
\begin{equation}
F_{R_o}(R_o\leq r\mid R_1, R_2) = \begin{cases} \frac{4r}{R_1 + R_2}, &\text{if}~  \frac{R_1}{2}\geq r \geq 0, \\
\frac{2r + R_1}{R_1 + R_2}, &\text{if}~   \frac{R_2}{2} \geq r > \frac{R_1}{2}, \\
1, &\text{if}~   r>\frac{R_2}{2}. \\
\end{cases}
\label{eq:Cond_CDF_1D}
\end{equation}
Now, we present the exact coverage probability of the Type I process in the following theorem.
\vspace{-.1cm}
\begin{thm}
The coverage probability of the Type I process in a 1-D Poisson cellular network is 
\begin{align}
\hspace{-.3cm}\mathtt{P_{I}^1}(\tau)=\int_{0}^{\infty} \hspace{-.25cm}\int_{0}^{r_2}\hspace{-.15cm}\dP{\sir>\tau\mid r_1, r_2}f_{R_1,R_2}(r_1, r_2) {\rm d}r_1 {\rm d}r_2,  
\label{eq:CovP_1d}
\end{align}
where $f_{R_1,R_2}(r_1,r_2)$ is given by \eqref{eq:DIST_R1R2}, 
\begin{align}
&\dP{\sir > \tau\mid r_1, r_2} = \int_{0}^{\frac{r_1}{2}}  {\cal L}_{\calI}(\tau r^{\alpha}\mid r_1-r, r_2+r)\frac{2}{r_1 + r_2} {\rm d}r\nonumber\\
&~~~~~~~~~~~+   \int_{0}^{\frac{r_2}{2}} {\cal L}_{\calI}(\tau r^{\alpha}\mid r_1+r, r_2-r)\frac{2}{r_1 + r_2} {\rm d}r,
\label{eq:CovP_Cond_1d}
\end{align}
\begin{align}
\hspace{-.25cm}~\text{and}~{\cal L}_{\calI}(s\mid u,v)=\frac{\exp\left(-\lambda \int_{u}^{\infty} \frac{s {\rm d}r}{r^{\alpha} + s} - \lambda \int_{v}^{\infty} \frac{s {\rm d}r}{r^{\alpha} + s} \right)}{(1 + s u^{-\alpha})(1 + s v^{-\alpha})}.
\label{eq:LT_Int_1D}
\end{align}
\end{thm}
\begin{IEEEproof}
Let ${x}_{l}$ and ${x}_{r}$ be the neighboring interfering BSs to the typical user at $y\in V_o$ in $\Psi_1$ and $\Psi_2$, respectively, where $\Psi_1=\Psi\cap\R^-$ and  $\Psi_2=\Psi\cap\R^+$. Let $\tilde{R}_1=|{x}_{l}-y|$ and $\tilde{R}_2=|{x}_{r}-y|$.
Thus, the aggregate interference can be written as $\calI=\calI_1+\calI_2$ where $\calI_1=h_{x_{l}}\tilde{R}_1^{-\alpha}+\sum_{x\in\Psi_1\setminus\{x_{l}\}} h_{x}|x-y|^{-\alpha}$ and $\calI_2=h_{x_{r}}\tilde{R}_2^{-\alpha}+\sum_{x\in\Psi_2\setminus\{x_{r}\}} h_{x}|x-y|^{-\alpha}$. Now, the Laplace transform (LT) of $\calI_1$ conditioned on $\tilde{R}_1$ is
\begin{align}
\calL_{\calI_1}(s\mid \tilde{R}_1)&=\E\left[e^{-sh_{x_{l}}\tilde{R}_1^{-\alpha}}\prod_{x\in\Psi_1\setminus\{x_{l}\}}e^{-sh_{x} |x-y|^{-\alpha}}\mid \tilde{R}_1\right]\nonumber\\
&\stackrel{(a)}{=}\E_{h}[e^{sh \tilde{R}_1^{-\alpha}}]\E\left[\prod_{x\in\Psi_1\setminus\{{x}_{l}\}}\E_{h}[e^{-sh|x-y|^{-\alpha}}]\mid\tilde{R}_1\right]\nonumber\\
&\stackrel{(b)}{=}\frac{1}{1+s \tilde{R}_1^{-\alpha}}\exp\left(-\lambda\int_{\tilde{R}_1}^\infty\frac{s{\rm d}r}{r^\alpha+s}\right),\nonumber
\end{align}
where (a) follows from the independence of the fading gains and (b) follows from the LT of an exponential r.v. and the probability generating functional ($\pgfl$) of the PPP \cite{Haenggi2013}. Similarly, we obtained LT of $\calI_2$ condition on $\tilde{R}_2$. Thus, the LT of aggregate interference conditioned on  $\tilde{R}_1=u$ and $\tilde{R}_2=v$ is given by \eqref{eq:LT_Int_1D}.
Now, conditioned on $R_1$ and $R_2$, the coverage probability becomes 
$$\dP{\sir > \tau\mid R_1, R_2}=\E_{R_o}\left[\calL_{\calI}(\tau r^\alpha\mid \tilde{R}_1,\tilde{R}_2)\mid R_1,R_2\right].$$ 
Finally, by deconditioning the above equation over the joint distribution of $R_1$ and $R_2$ given in \eqref{eq:DIST_R1R2}, we obtain \eqref{eq:CovP_1d}. 
\end{IEEEproof}
From the above analysis, it is evident that the exact analysis of the Type I process requires conditioning on the locations of {\em all the neighboring} BSs around $V_o$. While this was manageable in 1-D, it becomes significantly more complicated for $d>1$, which prevents an exact analysis. In the next subsection, we present a new approximation that leads to a tight characterization of the Type I user performance for $d=2$. 
\vspace{-.2cm}
\subsection{Approximate Coverage Analysis for $d=2$}
\label{sec:CovP_Analysis_2d}
The coverage analysis requires the joint distribution of the distances $\|\mathbf{x}-\mathbf{y}\|$, $\mathbf{x}\in\Psi$, and the link distance $R_o=\|\mathbf{y}\|$ of the typical user at $\mathbf{y}\sim U(V_o)$. Thus, we first discuss the distribution of $R_o$ and then approximate the point process of interferers $\Psi$ conditioned on $R_o$. Finally, using these distributions, we present the approximate coverage analysis.
\subsubsection{Approximation of the link distance distribution}
\label{subsec:CDF_Ro}
In \cite{PraPriHar}, we derived an exact expression for the distribution of $R_o$ which involves multiple integrals. Therein, we also derived a closed-form expression to approximate the $\cdf$ of $R_o$ which is 
\begin{equation}
F_{R_o}(r)\approx 1-\exp\left(-\pi\rho_o\lambda r^2\right),~\text{for}~r\geq 0,
\label{eq:CDF_Ro}
\end{equation} 
where $\rho_o=\frac{9}{7}$ is the correction factor (CF), which corresponds to the ratio of the mean volumes of $\tilde{V}_o$ and  $V_o$. 
 \begin{figure*}[h]
 \centering\vspace{-.2cm}
 \includegraphics[width=.32\textwidth]{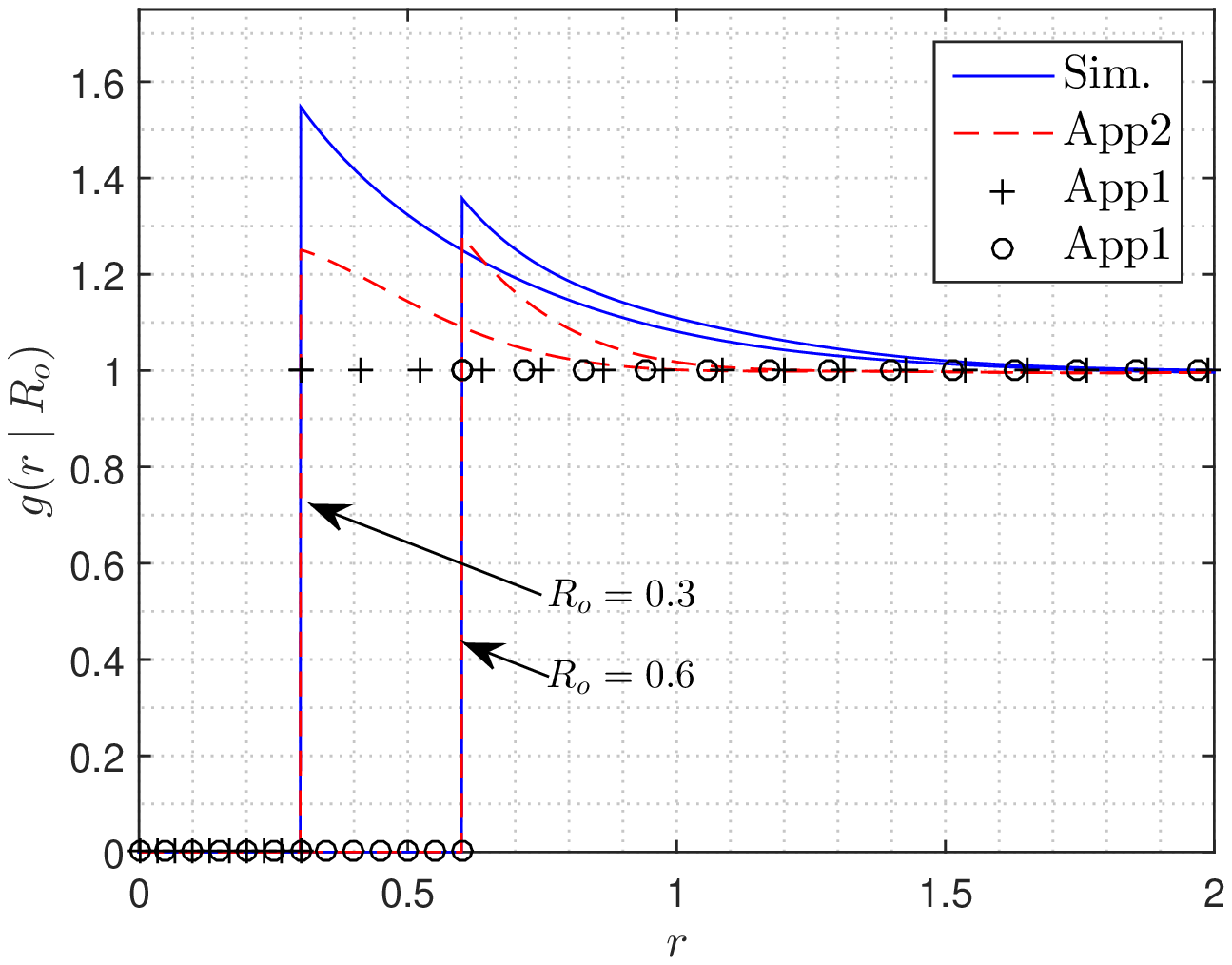}
\includegraphics[width=.32\textwidth]{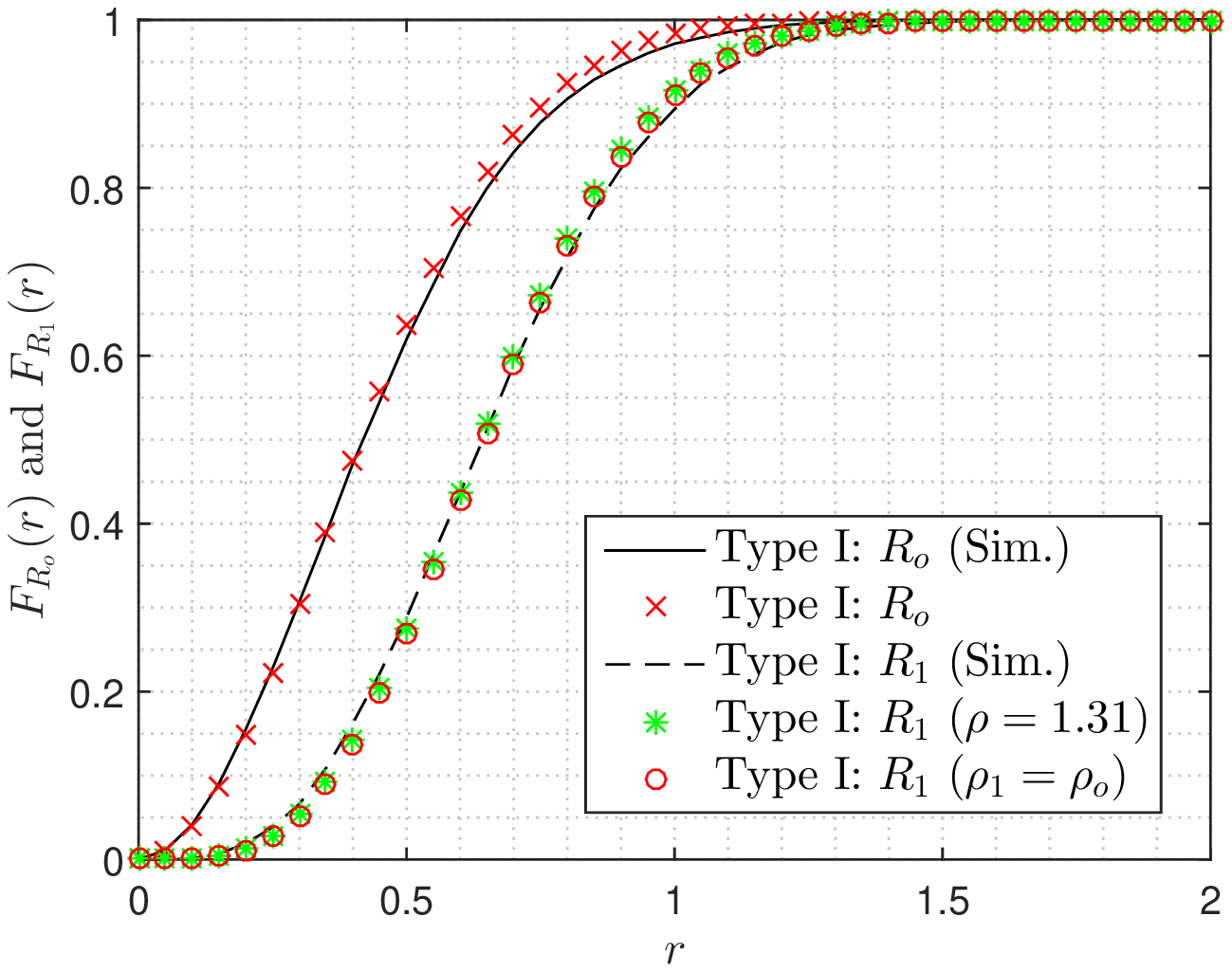}
\includegraphics[width=.32\textwidth]{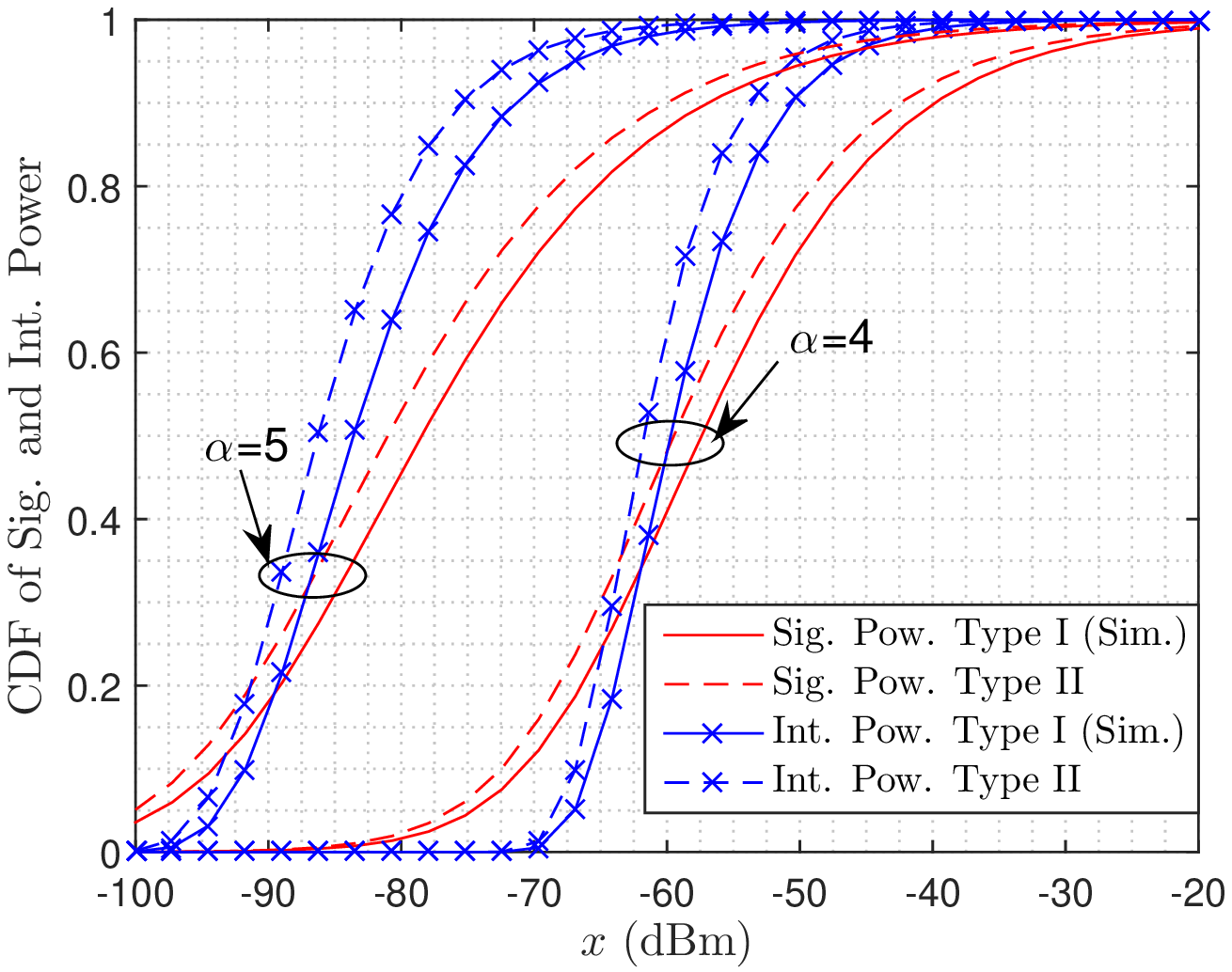}
\label{fig:CovP_TypCell}\vspace{-.25cm}
\caption{{\em Left:} The $\pcf$ of the point process of interferers observed by a user, from the Type I process, having link distance $R_o\in\{0.3,0.6\}$. {\em Middle:} Distributions of distances to the serving and dominant interfering BSs from the typical user of the Type I for $\lambda=1$. {\em Right:} CDF of desired signal and interference powers received by the typical users of the Type I and II processes for $\lambda=10^{-5}$ and $P=30$ dBm.}
\label{fig:CDF_Ro_R1_pcf}\vspace{-.3cm}
\end{figure*}
\subsubsection{Approximation of the point process of interferers $\Psi$}
\label{subsec:PP_PihI}
To understand the statistics of the point process of interferers observed by the typical user at $\mathbf{y}\in V_o$, we analyze the pair correlation function ($\pcf$) of $\Psi=\Phi\setminus\{o\}$ with reference to  $\mathbf{y}\in V_o$ which is \cite{Haenggi2013} 
\begin{equation}
g(r\mid R_o) = \frac{1}{2\pi r}\frac{{\rm d} K(r\mid R_o)}{{\rm d}r},~\text{for}~r>R_o,\nonumber
\end{equation}
where $R_o=\|\mathbf{y}\|$, $K(r\mid R_o)=\E[\Psi(\calB_\mathbf{y}(r))\mid R_o]$ is Ripley's $K$ function given $R_o$ and $\calB_\mathbf{y}(r)$ is the disk of radius $r$ centered at $\mathbf{y}$. Fig. \ref{fig:CDF_Ro_R1_pcf} (Left) shows the simulated user-interfering BS $\pcf$ conditioned on $R_o$. From the figure, it is easy to interpret that the point process of interferers exhibits a {\em clustering effect} at distances slightly larger than $R_o$ and complete spatial randomness for $r\gg R_o$.   The exact characterization of such point process is complex because of the correlation in the points (in $\Psi$) that form the boundaries of $V_o$ (as seen by the  typical user at $\mathbf{y}\in V_o$). Therefore, in order to accurately evaluate the interference received by the typical user, we need to carefully approximate the point process of interferers as seen by the typical user. 
 
 A natural candidate for the approximation is homogeneous PPP of density $\lambda$ outside of $\mathcal{B}_\mathbf{y}(R_o)$ \cite{Martin2017_Meta}. Henceforth, we refer to this approximation as $\mathtt{App1}$. $\mathtt{App1}$ ignores the clustering effect (see Fig. \ref{fig:CDF_Ro_R1_pcf} (Left)) and thus underestimates the interference. Therefore, in order to capture the effect of clustering to some degree, we explicitly consider the interference from the {\em dominant interferer} at distance $R_1=\arg\min_{\mathbf{x}\in\Psi}\|\mathbf{x}-\mathbf{y}\|$ and approximate the  point process of interferers with homogeneous PPP of density $\lambda$ outside $\mathcal{B}_{\mathbf{y}}(R_1)$.  We call this approximation $\mathtt{App2}$. Now, the crucial part is to obtain the distribution of $R_1$. 
Given the complexity of the analysis of r.v. $R_o$ \cite{PraPriHar}, it is reasonable to deduce that the exact characterization of the distribution of $R_1$ is equally, if not more, challenging. Thus, we obtain an approximate distribution of $R_1$ as follows.  

The $\cdf$ of $R_o$, given in \eqref{eq:CDF_Ro}, is the same as the contact distribution of PPP with density $\rho_o\lambda$. Therefore, using this and the argument of clustering discussed above, the $\cdf$ of $R_i$ (distance to $i$-th closed point in $\Psi$ from the user at $\mathbf{y}\in V_o$) can be approximated by inserting an appropriate CF $\rho_i$ to the $\cdf$ of $(i+1)$-th closest point to the origin in the PPP. From $g(r\mid R_o)\downarrow 1$ as $r\to\infty$, we have $\rho_i\to 1$ as $i\to \infty$. While $\rho_1=1.31$ gives the best fit for the empirical $\cdf$ of $R_1$, we approximate $\rho_1$ by $\rho_o$ for simplicity. Now, the $\cdf$ of $R_1$ conditioned on $R_o$ can be approximated as \cite{Moltchanov2012}
\begin{equation}
F_{R_1}(v\mid R_o)=1-\exp\left(-\pi\lambda\rho_o(v^2-R_o^2)\right)~\text{for}~v\geq R_o,
\label{eq:CDF_R1Ro}
\end{equation} 
and thus the approximated marginal $\cdf$ of $R_1$ becomes
\begin{align}
\hspace{-.2cm}F_{R_1}(v)=1-(\pi\lambda\rho_o v^2 + 1)\exp\left(-\pi\lambda\rho_o v^2 \right) ~\text{for}~ v\geq 0.
\label{eq:CDF_R1}
\end{align}
Fig. \ref{fig:CDF_Ro_R1_pcf} (Middle) shows the accuracy of the approximated CDFs of $R_o$ and $R_1$ given in \eqref{eq:CDF_Ro} and \eqref{eq:CDF_R1}.
Fig. \ref{fig:CDF_Ro_R1_pcf}(left) shows that App2 provides a slightly pessimistic estimate for the $\pcf$ because of which it will slightly underestimate the interference power.
\subsubsection{Coverage Probability}
\label{subsec:CoVP}
Now, we derive the coverage probability of the Type I process using the distribution of link distance $R_o$, given in \eqref{eq:CDF_Ro}, and the approximated point process of interferers $\mathtt{App2}$, discussed in Subsection \ref{subsec:PP_PihI}, in the following theorem.
\begin{thm}
The coverage probability of the Type I process in a 2-D Poisson cellular network can be approximated as 
\begin{align}
 \mathtt{P_{I}^2}(\tau)=\rho_o^2\tau^{-\delta}\int_{0}^{\tau^\delta}\frac{(\tilde\beta(t)+\rho_o)^{-2}}{1+t^{\frac{1}{\delta}}}{\rm d}t, 
 \label{eq:CovP}
 \end{align}
 where $\tilde\beta(t)=t\int_{t^{-1}}^{\infty}\frac{1}{1+u^{\frac{1}{\delta}}}{\rm d}u$ and $\delta=\frac{2}{\alpha}$.
\end{thm}
\begin{IEEEproof}
Let $\tilde{\mathbf{x}}$ be the dominant interfering BS such that $\|\tilde{\mathbf{x}}-\mathbf{y}\|=R_1$. We write the interference received by the user with link distance $R_o$ as $\mathcal{I}(R_o)=h_{\tilde{\mathbf{x}}}R_1^{-\alpha} + \mathcal{I}(\tilde{\Psi})$ where $\tilde{\Psi}=\Psi\setminus\{\tilde{\mathbf{x}}\}$ and $\mathcal{I}(\tilde{\Psi})=\sum_{\mathbf{x}\in\tilde{\Psi}}h_{\mathbf{x}}\|\mathbf{x}-\mathbf{y}\|^{-\alpha}$.
Thus, the coverage probability conditioned on $R_o$ and $R_1$ becomes
\begin{align}
\mathtt{P_{I}^2}&(\tau\mid R_o,R_1) = \P\left[h_o>\tau R_o^\alpha\mathcal{I}(R_o)\right]\nonumber\\
&=\calL_{h_{\tilde{\mathbf{x}}}}\left(\tau(R_o/R_1)^\alpha\right)\calL_{\calI(\tilde{\Psi})}\left(\tau R_o^\alpha\mid R_1\right).
\label{eq:CovP_Cond}
\end{align}
Now, the LT of $\calI(\tilde{\Psi})$ at $\tau R_o^{\alpha}$ for given $R_1$ can be obtained as
\begin{align}
&\calL_{\calI(\tilde{\Psi})}(\tau R_o^\alpha\mid R_1)\stackrel{(a)}{=}\E_{\tilde{\Psi}}\prod_{\mathbf{x}\in\tilde{\Psi}}\frac{1}{1+\tau R_o^{\alpha}\|\mathbf{x}-\mathbf{y}\|^{-\alpha}}\nonumber\\
&\stackrel{(b)}{=}\exp\left(-\lambda\int_{\R^2\setminus\calB_\mathbf{y}(R_1)}\frac{1}{1+\tau^{-1} R_o^{-\alpha}\|\mathbf{x}-\mathbf{y}\|^{\alpha}}{\rm d}\mathbf{x}\right)\nonumber\\
&\stackrel{(c)}{=}\exp\left(-2\pi\lambda R_o^2\tau^{\delta}\int_{\tau^{-\delta}(R_1/R_o)^2}^{\infty}\frac{1}{1+u^{\frac{1}{\delta}}}{\rm d}u\right),
\label{eq:LT}
\end{align}
where (a) follows due to the independence of the power fading gains and the LT of an exponential r.v., (b) follows using the $\mathtt{App2}$ and the $\pgfl$ of the PPP \cite{Haenggi2013}, and (c) follows through the substitution of $\mathbf{x}-\mathbf{y}=\mathbf{z}$ and then using Cartesian-to-polar coordinate conversion. 
Now, substituting \eqref{eq:LT} along with $\calL_{h_{\tilde{\mathbf{x}}}}\left(\tau(R_o/R_1)^\alpha\right)=\frac{1}{1+\tau(R_o/R_1)^\alpha}$ in \eqref{eq:CovP_Cond} and futher taking expectation over $R_1$ and $R_o$ yields the coverage probability as $\mathtt{P_{I}^2}(\tau)=$
\begin{align}
 \left(2\pi\lambda\rho_o\right)^2\int_{0}^{\infty}\int_{r}^{\infty}\frac{\exp\left(-\pi\lambda r^2\beta(\tau,r,v)-\pi\lambda\rho_o v^2\right)}{1+\tau(r/v)^\alpha}v{\rm d}v r{\rm d}r,\nonumber
  \end{align}
 where $\beta(\tau,r,v)=\tau^{\delta}\int_{\tau^{-\delta}(\frac{v}{r})^2}^{\infty}\frac{1}{1+u^{1/{\delta}}}{\rm d}u$. Now, by interchanging the order of the integrals  and further simplification, we obtain \eqref{eq:CovP}. This completes the proof.
\end{IEEEproof}\vspace{-.15cm}
\section{Numerical Results and Discussion}
Fig. \ref{fig:CDF_Ro_R1_pcf} (Right) shows that the desired signal power and interference power received by the typical users of the Type I and the Type II processes are significantly different (by 2-3 dB). Given the fundamental differences in the constructions of these two processes, this observation is not surprising. Besides, in Fig. \ref{fig:CovP_TypCell} we note that the coverage probabilities for the two processes are fairly similar, especially for higher values of $\alpha$. This is mainly because the desired signal power and the interference from a few dominant interfering BSs scale up by almost the same factors in the two cases (note that $\rho_1 \approx \rho_o$) and a few neighboring interfering BSs dominate the aggregate interference for higher values of $\alpha$.  A key point to note here is that the fact that the coverage probabilities are similar in the two models does not imply that the other performance measures will also be close. Finally, note that $\mathtt{App2}$ results in a slightly higher coverage probability since it slightly underestimates the pcf of the point process of interferers (refer Fig. \ref{fig:CDF_Ro_R1_pcf}(left)).
\vspace{-.35cm} 
\section{Conclusion}
In this letter, we have revisited the downlink analysis of cellular networks by arguing that the typical user analysis in the popular approach of considering a stationary and independent user point process results in the analysis of a Crofton cell, which is bigger on average than the typical cell. In order to characterize the performance of the typical cell, we consider a recent construction in which each cell is assumed to contain a single user distributed uniformly at random independently of the other cells. After highlighting the key analytical challenges in characterizing the typical cell performance in this case, we provide a remarkably accurate approximation that facilitates the general analysis of the typical cell in Poisson cellular networks. Even though the downlink coverage for the two cases is similar, we show that the other metrics, such as the received desired power and interference power, may exhibit significant differences, thus necessitating the need for a careful analysis of the typical cell. Although this letter was focused on the downlink coverage, the underlying characterization of the interference field can be used for the analysis of other key performance metrics as well. 
\begin{figure}[h]
 \centering\vspace{-.35cm}
\includegraphics[width=.45\textwidth]{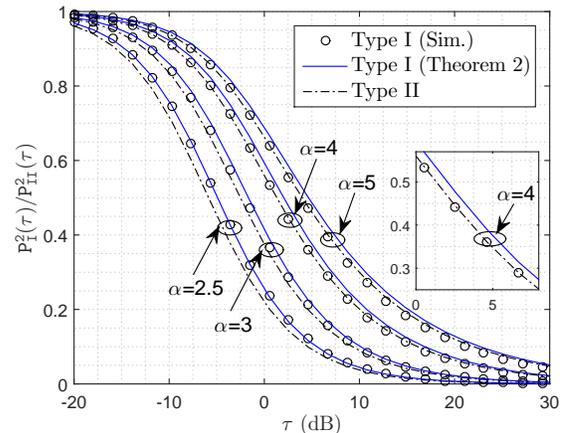}\vspace{-.35cm}
\caption{Coverage probability of the Type I and the Type II processes.}
\label{fig:CovP_TypCell}
\end{figure}
\vspace{-.4cm}

\end{document}